# Standing on the Shoulders of Their Peers: Success Factors for Massive Cooperation Among Children Creating Open Source Animations and Games on Their Smartphones

Tobias Gritschacher
Wikimedia Deutschland e.V.
Eisenacher Str. 2
10777 Berlin, Germany
tobias@gritschacher.eu

Wolfgang Slany
Institute for Software Technology
Graz University of Technology
Inffeldgasse 16b, 8010 Graz, Austria
wolfgang.slany@tugraz.at

## ABSTRACT
We developed a website for kids where they can share new as well as remixed animations and games, e.g., interactive music videos, which they created on their smartphones or tablets using a visual "LEGO-style" programming environment called Catroid. Online communities for children like our website have unique requirements, and keeping the commitment of kids on a high level is a continuous challenge. For instance, one key motivator for kids is the ability to entertain their friends. Another success factor is the ability to learn from and cooperate with other children. In this short position paper we attempt at identifying the requirements for the success of such an online community, both from the point of view of the kids as well as of their parents, and at finding ways to make it attractive for both.

## Categories and Subject Descriptors
H.5.m [**Information interfaces and presentation**]: Miscellaneous.

## 1. INTRODUCTION
The use of mobile phones and tablets is spreading rapidly over the world. This trend is not limited to adults. In industrial countries a growing number of children have their own mobile phone. E.g., already in 2010 more than 50% of German children aged between 6 and 13 years had their own mobile phone. Between the age of 12 and 13 years it was 90% [5]. Similar results can be seen in Asian countries like Japan, Korea, China or India [1]. An increasing number of children have access to a mobile internet connection allowing them to be online whenever and wherever they want. In Japan almost 70% of the children with mobile phones use them to access the Internet. In China and Korea the percentage is more than 30% [1] and in Germany 15% of the kids have a mobile internet connection [5]. These percentages are increasing quickly due to decreasing costs and the availability of flat rates. Other popular activities are making calls, writing messages, taking photos, and recording and exchanging videos.

The trend reflected in these statistics shows that the number of potential users of Catroid[1], our visual "LEGO-style" programming environment for smartphones and tablets, and of its online community website[2] for children, grows quickly and will rise even more in the near future.

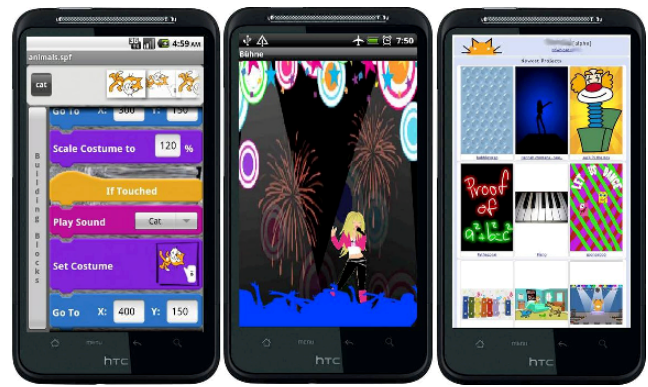

**Figure 1.** *Left*: **A Catroid project in development on a phone.** *Center*: **Execution of a Hannah Montana interactive music video animation using Catroid and created by children (remixed from an original Scratch project[3]).** *Right*: **Catroid projects on the community website.**

Our Catroid programming environment makes it easy for kids from the age of 8 to create interactive content such as multimedia animations or games directly on their smartphones without requiring previous knowledge about software development. The aim of our software is to empower children to gather their first programming experiences in a playful way. This paper focuses only on online community websites such as the one for Catroid projects that foster the cooperation between children and support them to get inspired and learn from others. The mobile visual programming environment Catroid itself is described in [9].

---
[1] http://code.google.com/p/catroid/
[2] http://catroid.org/
[3] http://scratch.mit.edu/projects/tyster/443306 available under the Creative Commons Attribution-ShareAlike License and created by the Scratch user known under the handle Tyster.

## 2. RELATED WORK

Our online community website is inspired by (but otherwise unrelated to) the thriving online community [3,7] for Scratch[4], the latter also being a visual programming language for children but in contrast to Catroid intended to be used on PCs or Laptops only. Catroid is inspired by Scratch but has many independent features related, among others, to the different possibilities and constraints of smartphones, e.g., position sensors, multi-touch screens, or small screen sizes, and has been developed independently of Scratch. Other online community websites supporting and encouraging the sharing of interactive animations and/or games created by kids include those associated with Nintendo's Wario Ware D.I.Y.[5] that runs on portable Nintendo DSi game consoles, Microsoft's Kodu[6] for the Xbox 360 game console, Flipnote Hatena[7] also for Nintendo DSi game consoles, and Game Maker[8] for PCs or Laptops. YouTube can also be seen as a platform to share user contributed multimedia content though it is not primarily oriented towards children and the contributed content cannot be made interactive.

## 3. GOALS

The goals of this paper are to identify unique requirements of online communities for children and to find success factors for such online communities of young creative people. To make such an online community website successful it is important that on the one hand it is highly attractive for kids and on the other hand does not cause any concerns from their parents. Thus we identify the unique requirements of such a site also from the parents' point of view.

## 4. SUITABILITY FOR CHILDREN, AND PARENTAL CONCERNS

One important issue when creating a website for children is how to handle inappropriate content. 16% of German children who were interviewed reported that they stumbled upon sites which were – in their own opinion – inappropriate for children. More than half of that inappropriate content was related to sex and about 20% to some form of violence. 11% have been in situations where they felt unpleasant or were afraid while visiting a website [5]. Looking at the parental concerns about their child's mobile phone usage shows that worries about inappropriate content is in the top three concerns together with overuse and cost of bills [1].

We hope to allay concerns about overuse by explaining how our software fosters both creativity as well as the mastering of programming skills and can be used as an educational tool for almost any curricular topic. To reduce the risk of exceeding the limit of one's data plan, which without an adequate flat rate could quickly lead to costly phone bills, we also clearly display the size of projects both before down- as well as before uploading them. Children are sensibilized to download volumes because of other content they download to their phones, but the increasing trend towards flat rates also contributes to alleviating associated risks.

To protect children from inappropriate content we take multiple measures. We immediately reject inappropriate content which is detectable automatically by filtering project titles and descriptions as well as comments and nicknames for cuss words. For inappropriate content not detected by the first method, e.g., inappropriate images, we make use of the power of social communities. We place a special report-as-inappropriate button on each project's details page giving users the possibility to actively take care of the community. Content reported as inappropriate stays hidden until an administrator has reviewed it. These measures are common and used on many websites which deal with user generated content. By having to give a reason why a project is reported as inappropriate this method *"also helps engage young people in thinking about their own moral reasoning"* [7].

## 5. MINIMIZE BARRIERS – MAXIMIZE COMMITMENT

Seeing the impact of the participation and getting reactions from others are crucial points when participating in an online community. This fact applies even more if the participants are children. Keeping the commitment to our community on a high level requires continuous effort. It begins with keeping the barriers of joining and using the community as low as possible. We aim at making this possible by minimizing the required steps for registration (participation without registration is also possible) and making the boundary between the app and the community fluid. These measures are even more important as we target mobile devices.

Another challenge is to encourage participation and get a high percentage of active content contributors. Zuckerman investigated the Israeli Scratch community and explained that participation can be increased by certain design features of the community website [10]. One of the design features we focus on is the arrangement of projects. For example a project not clearly showing up on the "newest projects" page in one's browser right after one has uploaded a project can discourage children to participate further. A reason why a project disappears too quickly could be simply because there are too many uploads at the same time. We aim at solving this by only showing a set of projects which are geographically related to the website visitor and by ensuring that one's own latest project is included in one's view of the webpage, for a certain time and until one accesses the page for the third time. Thus, e.g., a child in Korea will not see projects from Russia when viewing the "newest projects" page, and a girl in Shanghai will see her newly uploaded animation at the top of other new projects from her region in the first few minutes after she uploaded it even though these other projects might be newer. This could in the future be extended to showing this project also preferentially for a longer period of time to other children that viewed projects of this user before, thus increasing the chances that local or remote friends of a contributor will be able to see the project. Alerting users of new projects from their friends would also support this.

The suggestions page for Scratch[9] where users can request features for future versions of the online community website is a good source to find out what the Scratch community likes or dislikes. One issue frequently pointed out is that projects often stay the same on the "what the community is loving, viewing, and remixing" pages which means that the projects featured there have a much higher chance to get viewed, loved, and remixed again.

---

[4] http://scratch.mit.edu/
[5] http://www.warioawarediy.com/
[6] http://www.ditii.com/2011/03/03/kodu-with-community-game-sharing-released/
[7] http://flipnote.hatena.com/
[8] http://www.yoyogames.com/
[9] http://suggest.scratch.mit.edu/

The projects are distributed such that few top-placed projects are viewed, loved, and remixed extremely often. In contrast the majority consists of the less known projects. We have to focus also on projects located in this long tail. One simple method to realize this is a special "unknown projects" page with some randomly selected less popular projects which are presented for a defined period of time. Another possibility is to select the projects for the "what the community is loving, viewing, and remixing" pages randomly from a greater set of projects with the size of the set depending on the overall number of projects. Another method to increase the visibility of projects that are not in the top categories is to use automatic recommendation techniques as known from, e.g., Amazon's web shop, for instance by linking from each project to other projects that children viewed who also viewed the current project. This also helps in matching children with their particular interests and boosts the value of the projects in the parts of the long tail that are far away from the most popular projects, especially in combination with a search feature allowing to find projects based on keywords in their title, description, and comments. Other methods are the possibility for children to compile their own lists that can be made public, much like user contributed lists of one's favorite books on the websites of Amazon or movie lists on the Internet Movie Database. Such behavior increases the number of projects that can easily be discovered by browsing, thus leading to more satisfied users and a greater participation rate.

## 6. STANDING ON THE SHOULDERS OF ONE'S PEERS – COLLABORATION AND REMIXING

Manovich is citing Barb Dybwad when he writes *"[…] the most interesting aspects of Web 2.0 are new tools that explore the continuum between the personal and the social, and tools that are endowed with a certain flexibility and modularity which enables collaborative remixability – a transformative process in which the information and media we've organized and shared can be recombined and built on to create new forms, concepts, ideas, mashups and services"* [4]. It is an aim of our work to encourage children not only to create their own programs but to understand, learn from, and build on work of their peers. We want to help them to understand that collaboration and building on other people's work can lead to better results. They are enticed and encouraged to remix and improve projects of others. All projects uploaded to our online community website are open source and published under a free software license. This concept of remixing is also a core idea behind Scratch with currently around 60,000 remixes uploaded per month[10] [7]. YouTube started to encourage the remixing of videos published under a creative commons license at the beginning of June 2011[11] which shows the growing support of the remixing concept in other major online communities. Nevertheless, it is a challenge to make young people understand the principles of remixing and open source. Research regarding the responses to remixing in the Scratch community state that a significant number of children react negatively on remixes of their work, and some of them complain about plagiarism [2]. Automatic attribution of remixes and visualization of project genealogies do help but are not enough: E.g., children in the Scratch community to a high degree prefer credit given to them in a non-automatic, explicit way by the author of a remix of their original project [8]. Following a suggestion of Monroy-Hernández et al. [6,8] and to increase the acceptance of remixing, we design our system in a way that invites explicit manual crediting by providing an additional field for credits when uploading a remixed project. Additionally we provide a wiki and a forum for discussing the issues behind remixing, plagiarism, copyright and copyleft, and free open source software. These can be accessed directly when uploading a new or a remixed project, with contextually differing explanations for both cases. All this information should as much as possible be available in the children's language. Our project currently support several languages, with speakers of English, Mandarin, Cantonese, Hindi, Arabian, Japanese, Russian, German, Turkish, French, Urdu, Rumanian, and Malaysian in the team, and we use the crowd-sourcing software Pootle[12] to allow, e.g., parents to correct and/or add translations and new languages.

## 7. FOSTERING INTERNATIONAL COLLABORATION

Especially when designing an international website for children, supporting multiple languages is extremely important. Compared to translating a static website text, it is quite a challenge to support different languages throughout all user generated text. We plan to use the Google Translate API to optionally offer an automatic translation of project titles, project descriptions, comments, as well as text used inside of projects. Additionally to that we plan to support children in communicating with their foreign peers by introducing a simple template-based system for comments similar to the system used on Club Penguin[13]. Through this method kids can build comments out of a set of templates of common phrases. These templates are available in each supported language, again editable and extendable by the crowd-sourcing part of our project, and hence the most popular type of comments can potentially be displayed in the children' own language. All these methods further foster the cooperation among kids as the number of children that can easily communicate with each other is scaled up to a much larger portion of the world's population.

## 8. INTEGRATING SOCIAL NETWORKS

Over the last years social networks have become increasingly popular among children. The percentage of German children between the age of 6 and 13 who are using an online community at least once a week has grown from 16% in 2008 to 43% in 2010 [5]. 65% of Japanese and 27% of Chinese children who have a mobile internet connection use it to read messages and updates on their friend's profiles on social network sites [1]. Integrating support for popular social networks in a community like ours creates great opportunities both for the kids as well as for the popularity of our community. One key motivator for children to participate is the ability to entertain other kids and see their reactions, e.g., through their online feedback comments. By giving them the chance to easily promote their new projects on their social network profile they can reach an even wider audience. At the same time our community gets more attention by kids who usually might not be initially attracted.

A note on legal aspects: Even though laws regarding the involvement of children under, e.g., the age of 13 may vary from country to country, a website with social network features needs to be aware about legal issues, must take appropriate steps to protect its members, and must allow parents or legal guardians to exercise their rights. Examples from well established social networks such as Nickelodeon's neopets[14] can serve as best-practice examples.

---

[10] http://stats.scratch.mit.edu/
[11] http://youtube-global.blogspot.com/2011/06/youtube-and-creative-commons-raising.html
[12] http://translate.sourceforge.net/wiki/pootle/
[13] http://www.clubpenguin.com/

## 9. RESULTS

In summary, we have tentatively identified and partly implemented into our community website the following factors we deemed critical for the success of such a community:

- Minimize barriers for joining and using the community.
- The website is available in the children's native languages.
- The explanation of remixing and open source concepts must be easily understandable by children.
- Security aspects and filtering inappropriate content.
- Avoiding parental concerns like phone overuse, high costs, and inappropriate content.
- Clearly trigger the reactions of others to the project's original author in order to foster motivation and participation, in particular regarding remixes.
- Make the discovery of other projects of one's areas of interests as easy and intuitive as possible.

It must be noted that clearly the popularity of such a website is strongly influenced by the attractiveness of the associated software with which one creates the animations and games that can then be shared on the website, and vice versa.

## 10. CONCLUSION

Today kids are increasingly online whenever and wherever they are – mainly communicating, playing, surfing, or consuming audio or video using their mobile phones. Inspired by Scratch we developed a tool that allows children to become creators of their own interactive mobile content instead of just being consumers. The main difference to other similar systems is the mobile character of our system that allows children to use our system almost at any time and place, with the added possibilities available through the use of built-in sensors and actuators such as gyro-sensors or GPS and the challenge posed by the small screens, the strong encouragement of remixing and sharing, and the free open source availability of our software. We believe that this allows us to empower children to effectively create, and share their projects with their peers.

The ultimate measure of success of our website is the rate of growth of the popularity of the combined system. This would allow us to achieve our vision of empowering all children of the world to master the skills required to express themselves naturally and easily through programming the devices they carry with them at all times. We believe that this is an important endeavor, both for socio-economical as well as for deeper philosophical reasons, as programming can be seen as an automation of human rational thinking.

## 11. ACKNOWLEDGMENTS

Our thanks to the team members and supporters of Catroid[15].

---

[14] http://www.neopets.com/

[15] http://code.google.com/p/catroid/wiki/Credits